# Quantum dynamics of ultra-cold atoms in the presence of virtual photons

**Ghasem Asadi Cordshooli, Mehdi Mirzaee,***
*Department of Physics, Faculty of Science,
Arak University, Arak 38156-8-8349, Iran*
*Corresponding author: m-mirzaee@araku.ac.ir

A single trapped ion interacting with laser light in a radiofrequency trap is considered by diagonalization of full Hamiltonian of the system in a suitable basis. The energies, eigenvectors, probabilities of finding the atom in the ground state, density matrixes and its entanglement are computed. The study repeated for the Hamiltonian under the rotated wave approximation in the same basis. The results compared in general and in the Lamb-Dick regime for ultra-cold atoms to reveal essential changes in the results.

## I. Introduction

Dynamics of a trapped atom can be manipulated and controlled to cool down its vibrational motion up to ground state energy through coherent interaction of suitably tunned lasers [1, 2]. This mechanism known as laser cooling has become a fundamental technology by which some basic quantum concepts have been studied [3, 4] and made it possible to engineer desired atomic states [5]. Utrafine spectroscopy [6], quantum information processing [7] and enhanced atomic clocks [8] are other advanced applications of ultracold atoms. Cold atoms also proposed as suitable candidates of memory units in the quantum computers [9]. Quantum dynamics of cold trapped ions have been studied and considered in the field of quantum computations and applications [10, 11].

Historically, the seminal work of T.W. Hänsch and A.L. Schawlow in 1975, can be considered as the first attempt of cooling atoms through irradiating by light [12] after which a variety of laser cooling methods have been developed experimentally and reported [13-20]. Theoretically, Interaction of free two level atoms with photons has been described by the Jaynes-Cummings model (JCM) [21] and widely studied under the rotation wave approximation (RWA) [22]. The JCM is the cornerstone of atom manipulation by laser in different arrangements. One dimentional quantum motion of trapped ions is described in the Lamb-Dicke limit (LDL), in which the atom is localized in a region smaller than the lasers wavelength [1, 23]. The quantum dynamics of trapped ions was studied under the RWA and LDL and physical quantities was calculated for infinitely large time limit [24].

Whilst the RWA has been widely used to describe the atom-photon interaction in the resonance conditions, some exceptions have been resulted in wrong descriptions. Detuning of the photon and the atom, creates splitting in the wave function if non-RWA terms be taken into account in the Hamiltonian [25]. It was numerically shown that application of the RWA tends to the presence of nontrivial Berry phase in the atom-photon interaction described by the JCM [26]. Increasing transition probability [27], failing the estimation of Landau-Zender probability [28] and false conclusions in the JCM [26] are other reports on the insufficiency of the RWA.

In this paper we adopt a suitable basis to diagonalize the Hamiltonian of a system including a two level atom interacting with laser light in a radiofrequency trap. Energy eigenvalues of the system will be discussed. The time evolution of a selected initial state and related time dependent density matrix will be calculated. At last the concurence as a measure of entanglement will be calculated. The photons included in counter-rotating terms are known as the virtual photons for which the statics and dynamics are studied in [29, 30]. All treatments will be repeated for the Hamiltonian in the RWA in the same basis to study the role of virtual photons in different aspects of the problem. The results will be compared to the exact solutions in the ultracold condition.

This paper is structured as follows. The full and approximate Hamiltonians of the system will be introduced in the section II. The section III devotes to the diagonalization of the Hamiltonians and calculate their time dependent states, probabilities, density matrixes and concurrences. The computational results will be compared and discussed in the section IV. Summarized conclutions will be given in the last section.

## II. Model

The Hamiltonian of a two level atom with the ground state $|g\rangle$ and the excited state $|e\rangle$ is $\widehat{H}_A = \hbar(\omega_g|g\rangle\langle g| + \omega_e|e\rangle\langle e|)$. Defining $\omega_A = \omega_e - \omega_g$ and rescaling the energy such that $-\hbar(\omega_e + \omega_g) = 0$, gives $\widehat{H}_A = \frac{\hbar\omega_A}{2}\widehat{\sigma}_z$ in which $\widehat{\sigma}_z = |e\rangle\langle e| - |g\rangle\langle g|$ as described in [1]. The atom is confined in a RF trap of the frequency ν. In terms of the creation and annihilation operators of the trap, the motional Hamiltonian of the atom can be written as $\hat{H}_M = \hbar\nu\hat{a}^\dagger\hat{a}$. An external laser light as a running wave of the frequency $\omega_L$ and the wave vector $k_L$, derives the dipole transitions between the ground and excited atomic states. The interaction of laser with the two level atom can be written as $\hbar\frac{\Omega}{2}(\hat{\sigma}_- + \hat{\sigma}_+)(e^{i(\omega_L t - k_L \hat{q})} + e^{-i(\omega_L t - k_L \hat{q})})$ in which Ω is the Rabi frequency and $\hat{q} = \frac{\eta}{k_L}(\hat{a}^\dagger + \hat{a})$ denotes the atoms displacement operator from its equilibrium position in the trap. $\eta = k_L(\frac{\hbar}{2m\nu})^{1/2} = 2\pi\Delta x_{rms}/\lambda_L$ is called the Lamb-Dicke parameter [22]. As $\Delta x_{rms}$ is the root mean square of the atoms oscillations around its equilibrium position in the trap, small values of η defines the ultra-cold atoms. In this way the full Hamiltonian of the system can be written as

$$\hat{H} = \hbar\nu\hat{a}^\dagger\hat{a} + \hbar\omega_A\hat{\sigma}_z + \hbar\frac{\Omega}{2}\hat{\sigma}_x(e^{i(\omega_L t - k_L \hat{q})} + e^{-i(\omega_L t - k_L \hat{q})}), \quad (1)$$

in which $\hat{\sigma}_x = \hat{\sigma}_- + \hat{\sigma}_+$. The Hamiltonian (1) in the RWA condition reduces to

$$\widehat{H'} = \nu\hat{a}^\dagger\hat{a} + \hbar\omega_A\hat{\sigma}_z$$
$$+\hbar\frac{\Omega}{2}(\hat{\sigma}_- e^{i(\omega_L t - k_L \hat{q})} + \hat{\sigma}_+ e^{-i(\omega_L t - k_L \hat{q})}), \quad (2)$$

as given in [22]. The full Hamiltonian (1) and the approximate Hamiltonian (2) will be treated analytically in the next section.

### III. Analysis
#### A. Energy eigenvalues

In this section, the eigenvalues of the full Hamiltonian (1) and its reduced form under the RWA given by (2), will be obtained through diagonilzation in a suitably selected basis. Toward this end we need to rotate the Hamiltonians around y axis, π/2 degrees. Using the rotation operator $e^{i\pi\hat{\sigma}_y/2}$, the full Hamiltonian converts to

$$\hat{H}_R = \hbar\nu\hat{a}^\dagger\hat{a} - \hbar\omega_A\hat{\sigma}_x + \hbar\frac{\Omega}{2}\hat{\sigma}_z(e^{i(\omega_L t - k_L \hat{q})} + e^{-i(\omega_L t - k_L \hat{q})}). \quad (3)$$

Applying the operator theorem $e^{\hat{A}+\hat{B}} = e^{\hat{A}}e^{\hat{B}}e^{-[\hat{A},\hat{B}]/2}$, [31], and the commutation relation $[\hat{a}, \hat{a}^\dagger] = 1$, the Hamiltonian (3) takes the form

$$\hat{H}_R = \hbar\nu\hat{a}^\dagger\hat{a} - \hbar\omega_A\hat{\sigma}_x$$
$$+\hbar\frac{\Omega}{2}e^{-\frac{\eta^2}{2}}\hat{\sigma}_z(e^{-i\eta\hat{a}^\dagger}e^{-i\eta\hat{a}}e^{i\omega_L t} + e^{i\eta\hat{a}^\dagger}e^{i\eta\hat{a}}e^{-i\omega_L t}). \quad (4)$$

In the orthonormal basis,

$$\{|\psi_+\rangle, |\psi_-\rangle\} = \frac{1}{\sqrt{2}}\{|e,\alpha\rangle + |g,-\alpha\rangle, |e,\alpha\rangle - |g,-\alpha\rangle\}, \quad (5)$$

nondiagonal terms of the Hamiltonian (4) can be calculated as $H_{12} = H_{21} = +\hbar\Omega e^{-\frac{\eta^2}{2}}\cos(2\eta\alpha)\cos(\omega_L t)$ that are equal to zero when $\omega_L t = k\pi + \frac{\pi}{2}$ in which $k = 0,1,2,....$ So the Hamiltonian (4) will be diagonalized in the discrete times

$$t_k = \frac{1}{\omega_L}(k\pi + \frac{\pi}{2}). \quad (6)$$

The diagonal matrix elements of the Hamiltonian (4) at the time steps (6) take the form

$$H_{11} = \hbar\nu\alpha^2 - \hbar\omega_A e^{-2\alpha^2} + 2(-1)^k \hbar\Omega e^{\frac{-\eta^2}{2}}\sin(2\eta\alpha), \quad (7)$$

$$H_{22} = \hbar\nu\alpha^2 + \hbar\omega_A e^{-2\alpha^2} + 2(-1)^k \hbar\Omega e^{\frac{-\eta^2}{2}}\sin(2\eta\alpha). \quad (8)$$

Since the rotation has no effect on the eigenvalues of an operator, the results (7) and (8) are also the energy eigenvalues of the full Hamiltonian (1). The first terms in (7) and (8) are the vibrational energies of the atom in the trap, the second terms are the internal atomic states energies in which the negative and positive signs relate to the graund and excited states, respectively. The last terms describe the interaction energies of the atom and laser in the trap.

Diagonalization process of the full Hamiltonian can be repeated for the approximate Hamiltonian (2). Again, operating with the rotation matrix $e^{i\pi\hat{\sigma}_y/2}$, and applying the operator theorem $e^{\hat{A}+\hat{B}} = e^{\hat{A}}e^{\hat{B}}e^{-[\hat{A},\hat{B}]/2}$, givs

$$\widehat{H'} = \hbar\nu\hat{a}^\dagger\hat{a} - \hbar\omega_A\hat{\sigma}_x +$$
$$\hbar\frac{\Omega}{2}e^{-\frac{\eta^2}{2}}(\hat{\sigma}_- e^{-i\eta\hat{a}^\dagger}e^{-i\eta\hat{a}}e^{i\omega_L t} + \hat{\sigma}_+ e^{i\eta\hat{a}^\dagger}e^{i\eta\hat{a}}e^{-i\omega_L t}). \quad (9)$$

In the orthonormal basis (5), nondiagonal elements of the Hamiltinian (9), $\pm i\hbar\frac{\Omega}{2}\sin(\omega_L t)e^{-\frac{\eta^2}{2}-2\alpha^2}$, will be vanished under the condition $\omega_L t = k\pi$, with $k = 0,1,2,....$ In this way, the Hamiltonian (9) diagonalizes in the discrete times

$$t'_k = \frac{k\pi}{\omega_L}, \quad (10)$$

with the diagonal elements

$$H'_{11} = \hbar\nu\alpha^2 - \hbar\omega_A e^{-2\alpha^2} + (-1)^k \hbar\frac{\Omega}{2}e^{-\frac{\eta^2}{2}-2\alpha^2} \quad (11)$$

$$H'_{22} = \hbar\nu\alpha^2 + \hbar\omega_A e^{-2\alpha^2} - (-1)^k \hbar\frac{\Omega}{2}e^{-\frac{\eta^2}{2}-2\alpha^2}. \quad (12)$$

The first and second terms are the vibrational energy of the atom and the internal atomic states energies, respectively. The last terms are the laser-atom interaction energies. The energy eigenvalues will be compared and discussed in the section IV from the laser cooling viewpoint.

### B. Time dependent states

We choose the initial state $|e, \alpha\rangle$ and span it in the basis (5), to obtain

$$|e, \alpha\rangle = \frac{\sqrt{2}}{2}[|\psi_+\rangle + |\psi_-\rangle], \quad (13)$$

We intend to obtain the time evolution of this initial state. As the successive diagonalization times given by (5) are separated by $\frac{\pi}{\omega_L}$ that is commonly of the small order $10^{-6}$ in the experimental results [1], the time can be considered as a continiouse variable, approximately. In this way, the time evolution of the initial state (13) can be obtained by operating with the time evolution operator $e^{-i\hat{H}t/\hbar}$. Considering the $|\psi_+\rangle$ and $|\psi_-\rangle$ as eigenvectors of the rotated full Hamiltonian (4) with the eigenvalus (7) and (8), respectively, results in

$$|\psi(t)\rangle = \frac{\sqrt{2}}{2}[e^{-iH_{11}t/\hbar}|\psi_+\rangle + e^{-iH_{22}t/\hbar}|\psi_-\rangle]. \quad (14)$$

Replacing $H_{11}$ and $H_{22}$ from (7) and (8), respectively and $|\psi_+\rangle$ and $|\psi_-\rangle$ from (5) into (14), yields

$$|\psi(t)\rangle = e^{-i\delta t}[\cos(\theta)|e, \alpha\rangle + \sin(\theta)|g, -\alpha\rangle], \quad (15)$$

in which we defined

$$\delta = \nu\alpha^2 + 2(-1)^k \Omega e^{-\frac{\eta^2}{2}} \sin(2\eta\alpha), \quad \theta = \omega_A t e^{-2\alpha^2}. \quad (16)$$

Operating the time dependent state (15) by the inverse rotation operator $e^{-i\pi\hat{\sigma}_y/2}$, represents it in none rotated system as

$$|\psi(t)\rangle = e^{-i\delta t}[\cos(\theta)|g, \alpha\rangle - i\sin(\theta)|e, -\alpha\rangle]. \quad (17)$$

In a simillar procedure, operating the initial state (13) by $e^{-i\hat{H}'t/\hbar}$ tends to

$$|\psi'(t)\rangle = \frac{\sqrt{2}}{2}[e^{-iH'_{11}t/\hbar}|\psi_+\rangle + e^{-iH'_{22}t/\hbar}|\psi_-\rangle]. \quad (18)$$

Replacing $|\psi_+\rangle$ and $|\psi_-\rangle$ from (5) and $H'_{11}$ and $H'_{22}$ from (11) and (12), respectively gives

$$|\psi'^{(t)}\rangle = e^{-i\delta' t}[\cos\theta'|e, \alpha\rangle + \sin\theta'|g, -\alpha\rangle], \quad (19)$$

in which

$$\delta' = -\nu\alpha^2, \quad \theta' = \omega_A t e^{-2\alpha^2} - \Omega t e^{-\frac{\eta^2}{2} - 2\alpha^2(-1)^k}. \quad (20)$$

Operating (19) by $e^{-i\pi\hat{\sigma}_y/2}$ converts it to

$$|\psi'(t)\rangle = e^{-i\delta' t}[\cos(\theta')|g, \alpha\rangle - i\sin(\theta')|e, -\alpha\rangle], \quad (21)$$

that is the state in nonrotated initial framework. The time dependent states (17) and (21) will be considered to obtain the density matrix and calculate the entanglement in the next section.

The probability of being in the ground state at some given time is defined as

$$P_g(t) = \langle\psi(t)||g\rangle\langle g|\otimes I_m|\psi(t)\rangle, \quad (22)$$

in which $I_m$ is the identity operator in the motional space of the atom [1]. For the time dependent state (15), the probability $P_g(t)$ takes the form

$$P_g(t) = \cos^2(\theta)\langle\alpha|\otimes I_m|\alpha\rangle, \quad (23)$$

that is

$$P_g(t) = \cos^2(\theta). \quad (24)$$

Similarly, from (19) it can be calculated that

$$P'_g(t) = \cos^2(\theta'). \quad (25)$$

Considering the definations (16) and (20) and using the diagonalization times given by (6) and (10), the probabilities (24) and (25) take the form

$$P_g(t) = \cos^2[\frac{\omega_A}{\omega_L}\left(k\pi + \frac{\pi}{2}\right)e^{-2\alpha^2}], \quad (26)$$

$$P'_g(t) = \cos^2[k\pi(\frac{\omega_A}{\omega_L}e^{-2\alpha^2} - \frac{\Omega}{\omega_L}e^{-\frac{\eta^2}{2}-2\alpha^2(-1)^k})]. \quad (27)$$

$P_g(t)$ is independent of the Lamb-Dicke parameter but $P_g'(t)$ relates to it. The factor $(-1)^k$ in (20) implies that two different $|\psi'(t)\rangle$ can be considered for which two $P'_g(t)$ is expected, as given by (27).

### C. Density matrixes and entanglement

The density matrix of time dependent states (17) and (21) can be obtained by $\rho(t) = |\psi(t)\rangle\langle\psi(t)|$ as the density operator of a pure state [32]. considering the basis (5), the density matrix of the state (17) can be calculated as

$$\rho(t) = \cos^2\theta|g,\alpha\rangle\langle g,\alpha| + \sin^2\theta|e,-\alpha\rangle\langle e,-\alpha| - \frac{i}{2}\sin(2\theta)[|e,-\alpha\rangle\langle g,\alpha| - |g,\alpha\rangle\langle e,-\alpha|]. \quad (28)$$

For which the relations $\text{Tr}(\rho(t)) = 1$ and $\rho(t)^2 = \rho(t)$ holds, as expected.

Similarity, for the time dependent state (21), one obtaines

$\rho'(t) = \cos^2\theta'|g,\alpha\rangle\langle g,\alpha| + \sin^2\theta'|e,-\alpha\rangle\langle e,-\alpha| - \frac{i}{2}\sin(2\theta')[|e,-\alpha\rangle\langle g,\alpha| - |g,\alpha\rangle\langle e,-\alpha|].$ (29)

As expected, the relations $\text{Tr}(\rho'(t)) = 1$ and $\rho'^2(t) = \rho'(t)$ holds.

Here we calculate the concurrence of the density matrixes (28) and (29), as a measure of the entanglement. The concurrence is defined as $\max\{\lambda_1 - \lambda_2 - \lambda_3 - \lambda_4, 0\}$ in which $\lambda_1 \geq \lambda_2 \geq \lambda_3 \geq \lambda_4$ are the square roots of the eigenvalues of $\varrho \equiv \rho(\sigma_y \otimes \sigma_y)\rho^*(\sigma_y \otimes \sigma_y)$ in which $\sigma_y$ is a Pauli matrix [33]. The concurrence is nonzero if and only if the systems are entangled and is equal to zero only for separable matrixes. For a maximally entangled state, the concurrence equals one.

The matrix elements of (28) are

$\rho_{11} = \sin^2(\theta)|-\alpha\rangle\langle-\alpha|,$

$\rho_{12} = \frac{-i}{2}\sin(2\theta)|-\alpha\rangle\langle\alpha|,$

$\rho_{21} = \frac{i}{2}\sin(2\theta)|\alpha\rangle\langle-\alpha|,$

$\rho_{22} = \cos^2(\theta)|\alpha\rangle\langle\alpha|.$ (30)

To represent $\rho(t)$ in an orthogonal basis, we use $|\alpha\rangle = |0\rangle$ and $|-\alpha\rangle = M|1\rangle + P|0\rangle$ for which $P = e^{-2\alpha^2}$ and $M = \sqrt{1 - e^{-4\alpha^2}}$. So the density matrix elements (30) take the form

$\rho_{11} = \sin^2(\theta)[M^2|1\rangle\langle1| + P^2|0\rangle\langle0| + PM(|0\rangle\langle1| + |1\rangle\langle0|)],$

$\rho_{12} = \frac{-i}{2}\sin(2\theta)[M|1\rangle\langle0| + P|0\rangle\langle0|],$

$\rho_{21} = \frac{i}{2}\sin(2\theta)[M|0\rangle\langle1| + P|0\rangle\langle0|],$

$\rho_{22} = \cos^2(\theta)|0\rangle\langle0|.$ (31)

The concurrence of the density matrix (28) will be obtained as

$C = \frac{1}{2}(1 - e^{-4\alpha^2})(1 - \cos(4\omega_A t e^{-2\alpha^2})).$ (32)

Similarly, for the density matrix (29) we obtain

$C' = \frac{1}{2}(1 - e^{-4\alpha^2})(1 - \cos(4\omega_A t e^{-2\alpha^2} - 4\Omega t e^{-\frac{\eta^2}{2} - 2\alpha^2(-1)^k})).$ (33)

The concurrences (32) and (33) will be discussed in the next section.

### IV. Discussions

The full Hamiltonian of atom-photon interaction in a radiofrequency trap and its approximate form under the RWA was considered analytically in the previous section. The results including diagonalization times, energy eigenvalues, time dependent states, probabilities and the entanglement of both Hamiltonians will be discussed and compared in this section.

Diagonalization times

Diagonalization of the rotated Hamiltonians (4) and (9) was restricted to discrete times (6) and (10), respectively. As a result, the RWA causes that the Hamiltonian be diagonalized $\frac{\pi}{2\omega_L}$ faster than full Hamiltonian. Discrete diagonalization times for both Hamiltonians have equal time steps given by $\frac{\pi}{\omega_L}$.

Energy eigenvalues

The vibrational energy ($\hbar\nu\alpha^2$) and internal states energies ($\pm\hbar\omega_A e^{-2\alpha^2}$) of the atom for full Hamiltonian given in the (7) and (8) are the same of approximate Hamiltonian given by (11) and (12). Taking into account the role of virtual photons, just affects the interaction energy

$E_{int} = 2\hbar\Omega e^{-\frac{\eta^2}{2}}\sin(2\eta\alpha)$ (34)

of the full Hamiltonian and

$E'_{int} = \frac{1}{2}\hbar\Omega e^{-\frac{\eta^2}{2} - 2\alpha^2}$ (35)

of the approximate Hamiltonian. Total energy of the atom splits by adding and subtracting $E_{int}$ and $E'_{int}$. The interaction energies (34) and (35) are plotted in the figure 1 as functions of $\eta$ and $\alpha$.

$E_{int}$, vanishes in the limiting case, $\eta \to 0$, and changes linearly in terms of $\alpha$ for small $\eta$ values and for large $\eta$ values it oscilates with $\alpha$. $E'_{int}$ tends to $\frac{1}{2}\hbar\Omega e^{-2\alpha^2}$ when $\eta \to 0$ and behaves as a Gaussian function of $\alpha$ for nonzero $\eta$ values.

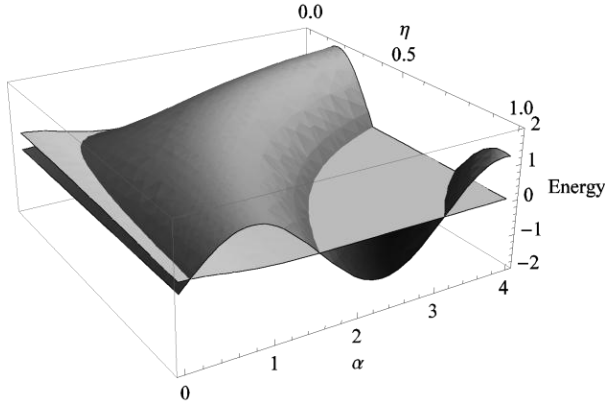

Fig. 1. Interaction energies of the full and approximate Hamiltonians. Dark surface shows $E_{int}$ given by (28) and light surface shows $E'_{int}$ given by (29) in the $\hbar\Omega$ unit.

Minimum energies of the systems described by the full and approximate Hamiltonians can be acquired for odd k values from (7) and (11), in which the greater interaction energy determines the smaller total energy of the system. The inequality $E_{int} - E'_{int} > 0$ determines some regions in the $\eta - \alpha$ surface where in more cooling is predicted by full Hamiltonian. These regions are shown in the figure 2 by gray color.

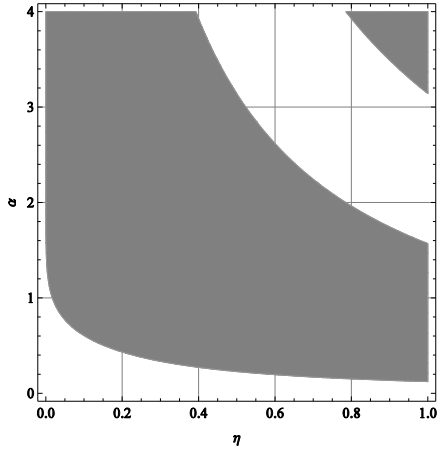

Fig. 2. Regions in $\eta - \alpha$ surface for which $E_{int} - E'_{int} > 0$.

The gray regions show $\eta$ and $\alpha$ values for which the energies of the exact solution are smaller than the energies of the RWA solutions and in the white regions, the energies of the full Hamiltonian are greater than the energies of the approximate Hamiltonian. We can say taking into account the virtual photons in the Hamiltonian, decreases the minimum energy of the system in the most of η-α area, implying more cooling of the atoms can be achieved.

Time dependent states and probabilities

Considering the initial state $|e, \alpha\rangle$ and assuming the continuity of the times of diagonalization, the time dependent states (15) and (19) was obtained for the full and approximate Hamiltonians, respectively. The similarly structured states differ in the phase factors, $\delta, \delta'$ and the angles $\theta, \theta'$ given by (16) and (20) in such a way that the Rabi frequency and Lamb-Dicke parameter take part in the phase factor for full Hamiltonian while they take part in the arguments of the trigonometric coefficients of the approximate Hamiltonian. Considering the frequencies $\omega_A$ and $\omega_L$ applied in the laser cooling experiments [1] that are of the order $10^6$, respectively, the probabilities given by (26) and (27), take the form

$$P_g(t) = \cos^2\left[\left(k\pi + \frac{\pi}{2}\right)e^{-2\alpha^2}\right] \qquad (36)$$

and

$$P'_g(t) = \cos^2[k\pi(e^{-2\alpha^2} - 10^{-3}e^{-\frac{\eta^2}{2}-2\alpha^2(-1)^k})] \qquad (37)$$

that are plotted in the figure 3 for $k = 0,1,2,3$. At $k = 0$, the probability of finding the atom in the ground state, as predicted by the full Hamiltonian, starts from zero and goes toward one by increasing α, while according to the approximate result (37), the probability is equal to one, for all α values.

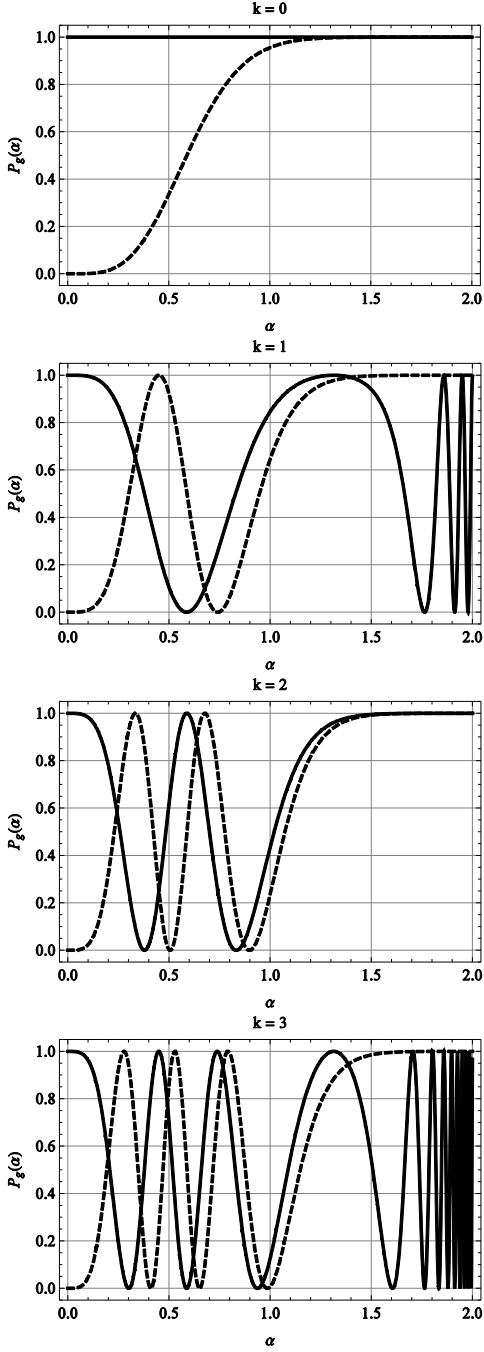

$\alpha \to 2$. The $P'_g$ shows two essential impediments. It is independent of $\alpha$ at initial diagonalization time and its behavior differs for consecutive even and odd times.

Entanglement

We consider the diagonalization times of the full and approximate Hamiltonians given by (6) and (10) to compare the entanglements, (28) and (29), respectively. According to [1], the Rabi frequency is of the order $10^3$, so we have $\frac{\Omega}{\omega_L} \sim 10^{-3}$ and $\frac{\omega_A}{\omega_L} \sim 1$ for both resonant and red sidebands. In this way

$$C = \frac{1}{2}(1 - e^{-4\alpha^2})(1 - \cos[4(k\pi + \frac{\pi}{2})e^{-2\alpha^2}]) \qquad (38)$$

from (32), and

$$C' = \frac{1}{2}(1 - e^{-4\alpha^2})(1 - \cos[4k\pi(e^{-2\alpha^2} - 10^{-3}e^{-\frac{\eta^2}{2} - 2\alpha^2(-1)^k})]) \qquad (39)$$

according to (33). There are some interesting differences in (38) and (39). At first, the concurrence is independent of the Lamb-Dick parameter when we work with the full Hamiltonian, while it takes part in the concurrence under the RWA. As shown in the figure 4, for $k = 0$, the entanglement is a nonzero function of $\alpha$ in the exact analysis while in the RWA viewpoint, the atomic vibrational states and the atomic internal state are disentangled.

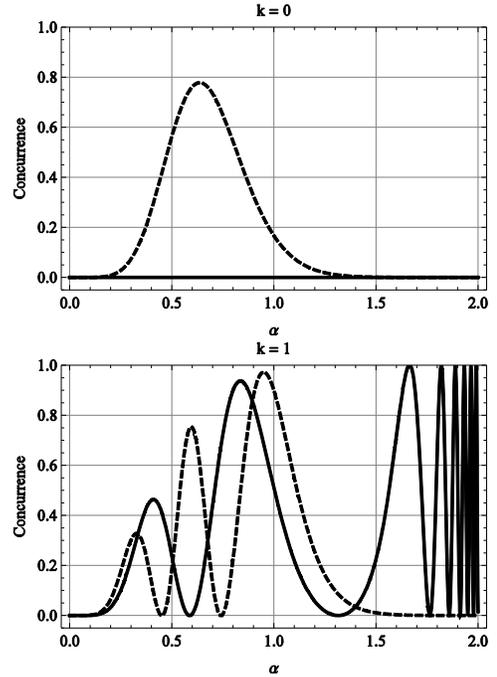

Fig. 3. The probabilities (36) and (37) of the atom to be in the ground state based on the full and approximate eigenvectors, respectively.

For $k > 0$ the probability $P_g$ has a general form as a function of $\alpha$. It starts from zero and oscillates faster for middle $\alpha$ values, as $k$ increases. The $P'_g$ behaves differently for even and odd $k > 0$, as a function of $\alpha$. For even $k$, it starts from zero and tends to one after some oscillations while for odd $k$ values, it oscillats with high frequency as

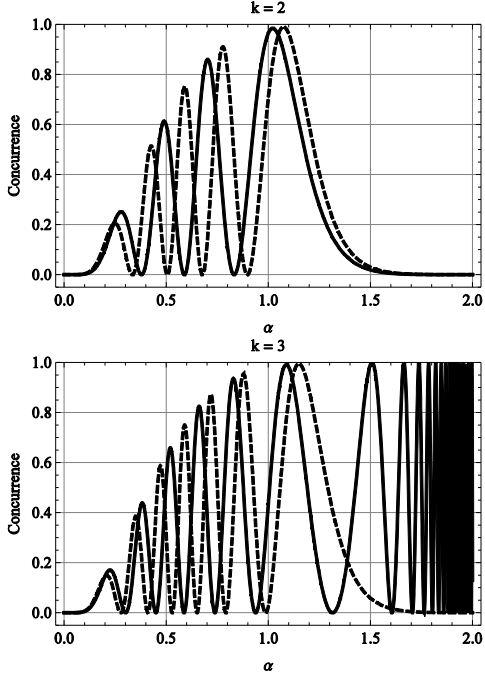

Fig. 4. The concurrence given by (38) and (39) as the entanglement criterions of the full and approximate density matrixes. Solid line indicates the concurrence of approximate density matrix (29) and dashed line shows the concurrence of the full density matrix (28).

For other even $k$ values, the entanglements are formally similar and differ only in the phase of cosine term, but for odd $k$ values, due to positive sign of $-2\alpha^2(-1)^k$ in the argument of cosine term, there is an essential difference in the concurrences. For $\alpha > 1.5$ the concurrence oscillates by a high frequency between zero and its maximum values.

## V. Conclusions

In this paper we considered a single atom interacting with laser light in a RF trap and studied the role of virtual photons on many essential parameters of the system including starting diagonalization time of the Hamiltonian, interaction energies of the system, minimum energy of the systems, probabilities of the atom to be in the ground state and the entanglements of density matrixes of the systems. The interaction energy obtained by diagonalization of the full Hamiltonian predicts more cooling of the atoms in the most area of $\eta - \alpha$ plane. The probability of finding the atom in the ground state as a function of $\alpha$ differs formally for successive times based on even and odd $k$ values. A similar behavior was observed for the entanglement in successive discrete times, while for the full Hamiltonian its general form is independent of time. The entanglement of density matrix of full Hamiltonian was independent of the Lamb-Dicke parameter, while it obtained as a function of this parameter in the RWA condition.

The results given in this paper, open new prospects for more accurate analytical considerations of atom-photon interactions, especially in different laser cooling mechanisms.

**References**

[1] Quantum dynamics of single trapped ions, D. Leibfried, R. Blatt, C. Monroe, and D. Wineland, Rev. Mod. Phys. 75, 281 (2003).

[2] Laser cooling of trapped ions, J. Eschner, G. Morigi, F. Schmidt-Kaler, R. Blatt, J. Opt. Soc. Am. B 20 1003 (2003).

[3] An experimental test of non-local realism, Simon Gröblacher, Tomasz Paterek, Rainer Kaltenbaek, Caslav Brukner, Marek Zukowski, Markus Aspelmeyer and Anton Zeilinger, Nature 446, 871-875 (2007).

[4] Experiment and the foundations of quantum physics, Anton Zeilinger, Rev. Mod. Phys. 71, S288 (1999).

[5] Engineering quantum pure states of a trapped cold ion beyond the Lamb-Dicke limit, L. F. Wei, Yu-xi Liu, and Franco Nori, Phys. Rev. A 70, 063801 (2004).

[6] Laser Cooling and Double Resonance Spectroscopy of Stored Ions, H.G. Dehmelt, Bull. Am. Phys. Soc. 20, 60 (1975).

[7] Electrodynamically trapped Yb+ ions for quantum information processing, Chr. Balzer, A. Braun, T. Hannemann, Chr. Paape, M. Ettler, W. Neuhauser, and Chr. Wunderlich, Phys. Rev. A 73, 041407(R) (2006).

[8] Trapped ions, laser cooling, and better clocks, Wineland D. J, Science. 226(4673), 395-400 (1984).

[9] Quantum Computations with Cold Trapped Ions, J. I. Cirac, P. Zoller, Pphys. Rev. lett, 74, 20 (1995).

[10] Quantum dynamics of cold trapped ions with application to quantum computation, D. F. V. James, Appl. Phys. B 66, 181-190 (1998).

[11] Quantum computation with two-level trapped cold ions beyond Lamb-Dicke limit, L. F. Wei, S. Y. Liu, X. L. Lei, Phys. Rev. A, 65, 062316 (2002).

[12] Cooling of gases by laser radiation, T.W. Hänsch, A.L. Schawlow, Optics Communications, 13, 1, 68-69 (1975).

[13] Microwave-Optical Double Resonance on a Single Laser-Cooled 171Yb+ Ion, V. Enders, Ph. Courteille, R. Huesmann, L. S. Ma, W. Neuhauser, R. Blatt and P. E. Toschek, Europhysics Letters, 24, 5, (1993).

[14] Laser cooling by spontaneous anti-Stokes scattering, N. Djeu, W. T. Whitney, Phy. Rev. Let, 46, 236 (1981).

[15] Resolved-Sideband Raman Cooling of a Bound Atom to the 3D Zero-Point Energy, C. Monroe, D. M. Meekhof, B. E. King, S. R. Jefferts, W. M. Itano, D. J. Wineland, and P. Gould, Phys. Rev. Lett. 75, 4011 (1995).

[16] Laser cooling of ions stored in harmonic and Penning traps, W. M. Itano, DJ Wineland, Phy. Rev. A (1982).

[17] Laser Cooling to the Zero-Point Energy of Motion, F. Diedrich, J. C. Bergquist, Wayne M. Itano, and D. J. Wineland, Phys. Rev. Lett. 62, 403 (1989).

[18] S Chu, L Hollberg, JE Bjorkholm, A Cable, A Ashkin, Three-dimensional viscous confinement and cooling of atoms by resonance radiation pressure, Phys. Rev. Lett. 55, 48, (1985).

[19] A "Schrödinger Cat" Superposition State of an Atom C. Monroe, D. M. Meekhof, B. E. King, D. J. Wineland, Science, 272, 5265, 1131-1136, (1996).

[20] Experimental Demonstration of Ground State Laser Cooling with Electromagnetically Induced Transparency, C. F. Roos, D. Leibfried, A. Mundt, F. Schmidt-Kaler, J. Eschner, and R. Blatt, Phys. Rev. Lett. 85, 5547 (2000).

[21] Comparison of quantum and semiclassicalradiation theories with application to the beam mase, E.T. Jaynes and F. W. Cummings, Proc.IEEE, 51(1) 89 (1963).

[22] Quantum optics, D. F. Walls, Gerard J. Milburn, (Springer 2008).

[23] Laser cooling of trapped ions in a standing wave, J. I. Cirac, R. Blatt, P. Zoller, W. D. Phillips, Phy. Rev. A, 46, 5 (1992).

[24] Measured quantum dynamics of a trapped ion, Lorenza Viola, Roberto Onofrio, Phys. Rev. A, 55, R3291(R) (1997).

[25] Optical Stern-Gerlach effect beyond the rotating-wave approximation, V. E. Lembessis, Phys. Rev. A 78, 043423 (2008).

[26] Absence of Vacuum Induced Berry Phases without the Rotating Wave Approximation in Cavity QED, J. Larson, Phys. Rev. Lett, 108, 033601 (2012).

[27] Amplitude spectroscopy of two coupled qubits, A. M. Satanin, M. V. Denisenko, Sahel Ashhab, and Franco Nori, Phys. Rev. B 85, 184524 (2012).

[28] Photon-assisted Landau-Zener transition: Role of coherent superposition states, Zhe Sun, Jian Ma, Xiaoguang Wang, and Franco Nori, Phys. Rev. A 86, 012107 (2012).

[29] Vacuum Rabi oscillation induced by virtual photons in the ultrastrong-coupling regime, C. K. Law, Phys. Rev. A 87, 045804 (2013).

[30] Atoms dressed and partially dressed by the zero-point fluctuations of the electromagnetic field, G. Compagnot, G. M. Palmat, R. Passantet and F. Persicott, 1105, J. phys. B At. Mol. Opt. Phys. 28, 1105 (1995).

[31] Quantum Mechanics, vol 1, A. Messiah, North-Holland Publishing Company, Amsterdam, (1961).

[32] Entangled systems: New Directions in Quantum Physics , Jurgen Audretsch, WILEY-VCH (2007).

[33] Entanglement of a Pair of Quantum Bits, S. Hill and W. K. Wootters, Phys. Rev. Lett. 78, 5022 (1997).